# Negative differential thermal conductance between Weyl semimetals nanoparticles through vacuum

Yasong Sun[1,2], Yang Hu[1,2,3], Kezhang Shi[4], Jihong Zhang[5], Dudong Feng[6], and Xiaohu Wu[3,*]

[1] *Basic Research Center, School of Power and Energy, Northwestern Polytechnical University, Xi'an 710072, Shaanxi, P. R. China*

[2] *Center of Computational Physics and Energy Science, Yangtze River Delta Research Institute of NPU, Northwestern Polytechnical University, Taicang 215400, Jiangsu, P. R. China*

[3] *Shandong Institute of Advanced Technology, Jinan 250100, Shandong, P. R. China*

[4] *Centre for Optical and Electromagnetic Research, National Engineering Research Center for Optical Instruments, Zhejiang University, Hangzhou 310058, P. R. China*

[5] *School of Electromechanical and Automotive Engineering, Yantai University, Yantai 264005, Shandong, P. R. China*

[6] *School of Mechanical Engineering, Purdue University, West Lafayette, IN 47907, USA*

*Email: xiaohu.wu@iat.cn

**Abstract**

In this work, the near-field radiative heat transfer (NFRHT) between two Weyl semimetal (WSM) nanoparticles (NPs) is investigated. The numerical results show that negative differential thermal conductance (NDTC) effect can be obtained in this system, i.e., when the temperature of the emitter is fixed, the heat flux does not decrease monotonically with the increase of the temperature of the receiver. Specifically, when the temperature of the emitter is 300 K, the heat flux is identical when the temperature of the receiver is 50 K or 280 K. The NDTC effect is attributed to the fact that the permittivity of the WSMs changes with the temperature. The coupling effects of polarizability of two WSM NPs have been further identified at different temperature to reveal the physical mechanism of the NDTC effect. In addition, the NFRHT between two Weyl WSM NPs can be greatly enhanced by exciting the localized plasmon and circular modes. This work indicates that the WSMs maybe promising candidate materials for manipulating NFRHT.

# 1. Introduction

Negative differential thermal conductance (NDTC) phenomenon was firstly observed from the negative differential resistance, and then it was extended to the field of radiation, which has broad application prospects [1-5]. For the radiative heat transfer between emitter and receiver with temperatures $T_1$ and $T_2$, respectively, the thermal conductance can be defined as $\left.(\partial P/\partial T_1)\right|_{T_2}$ and $\left.(-\partial P/\partial T_2)\right|_{T_1}$, where $P$ is heat flux from emitter to receiver [5]. In the radiative heat transfer between two blackbodies, the thermal conductance is always positive. However, the thermal conductance can be negative when the permittivity of the emitter or the receiver changes with the temperature. In addition, in the near-field radiative heat transfer (NFRHT) between two objects, and when the permittivity of one object changes with the temperature, the NDTC was more obvious due to the contribution of evanescent waves [6-30].

Weyl semimetals (WSMs) have attracted much attention in recent years, because they can inherently break time-reversal symmetry and exhibit extremely large nonreciprocity in the mid-infrared frequencies [31, 32]. Besides, the chemical potential of WSMs can be changed by controlling the temperature, thus the permittivity of WSMs will change with the temperature [33]. Zhao et al. investigated the nonreciprocal radiation in WSM gratings and they also pointed out that radiative properties of radiation could be flexibly tuned by changing the temperature [34]. Therefore, it is possible to realize NDTC with WSMs. Guo et al. have studied the NFRHT between three WSM nanoparticles (NPs) [35]. They found that the radiative thermal router could be realized. However, the change of the temperature was not fully investigated in their work and the NDTC was not discussed.

In this work, we investigate the NFRHT between two WSM NPs. Particularly, the heat flux as a function of the temperature of the receiver is comprehensively analyzed. When the temperature of the emitter is fixed, the heat flux does not decrease monotonically with the increase of the temperature of the receiver, indicating the NDTC effect. The physical mechanism behind this effect is that the permittivity of the WSM changes with the temperature. Besides, it is found that the excitation of localized plasmon modes can enhance the NFRHT when the value of momentum-separation $b$ (the momentum-separation means the separation of a pair of Weyl nodes in momentum space) is zero. However, localized plasmon and circular modes can be excited when the value of momentum-separation $b$ is nonzero. This work shows that the WSMs are promising materials to manipulate the NFRHT.

## 2. Modeling and calculation

Figure 1 shows the schematic illustration of a NDTC system based on NFRHT between WSM NPs. Here, two particles are described as point sources with radius much smaller than the thermal wavelength. When the particle spacing is greater than three times the radius, the dipole approximation can be well applied [43, 44]. In the calculation, the temperatures of the emitter and receiver are $T_1$ and $T_2$, respectively. The coordinate of emitter and receiver are $\mathbf{r}_1 = (a\ b\ c)$, and $\mathbf{r}_2 = (a+d\ b\ c)$, respectively. The distance between the emitter and receiver is $d=200$ nm, and the radius of two particles is $R=10$ nm.

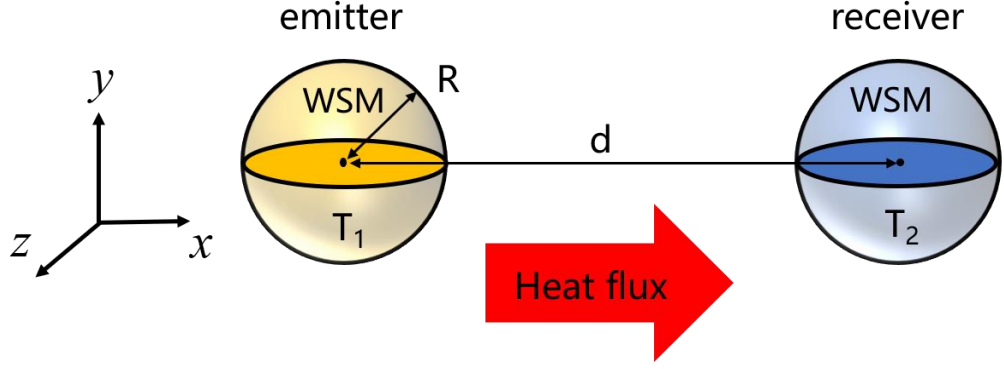

Fig. 1 Schematic of a proposed NDTC system based on NFRHT between WSM NPs. The two particles with radius *R* are placed along the *x*-axis at a gap distance of *d*. The temperatures of the emitter and the receiver are $T_1$ and $T_2$, respectively.

In WSM, the time-reversal symmetry could be broken when splitting a Dirac point into a pair of Weyl nodes with opposite chirality. Each pair of Weyl nodes is separated in momentum space by wavevector **b**. The presence of Weyl nodes changes the electromagnetic response, and the displacement electric field for WSM can be written as [36]

$$\mathbf{D} = \varepsilon_d \mathbf{E} + \frac{ie^2}{4\pi^2 \hbar \omega}\left(-2b_0 \mathbf{B} + 2\mathbf{b} \times \mathbf{E}\right) \quad (1)$$

Here, $\varepsilon_d$ is the permittivity of the corresponding Dirac semimetal, **E** is the electric field, and **B** is the magnetic flux density. In the absence of an external magnetic field, Dirac semimetal is generally assumed to be isotropic. Below, we assume that the diagonal element of the permittivity tensor is the same $\varepsilon_d$. The first (-2$b_0$**B**) and the second (2**b**×**E**) terms in the parentheses of Eq. (1) describe the chiral magnetic effect and the anomalous Hall effect, respectively. In this paper, we consider only materials where the Weyl nodes have the same energy (i.e., $b_0 = 0$). The momentum-separation **b** of the Weyl nodes is an axial vector that acts similar to an internal magnetic field, and we choose the coordinates with **b** along the *x*-direction: (i.e., **b**=b**x**). With the above considerations, the permittivity tensor of the WSM becomes

$$\boldsymbol{\varepsilon} = \begin{bmatrix} \varepsilon_d & 0 & 0 \\ 0 & \varepsilon_d & i\varepsilon_a \\ 0 & -i\varepsilon_a & \varepsilon_d \end{bmatrix} \quad (2)$$

where

$$\varepsilon_a = \frac{be^2}{2\pi^2 \varepsilon_0 \hbar \omega} \quad (3)$$

When $b \neq 0$, $\varepsilon_a$ is nonzero, therefore $\varepsilon$ is asymmetric and could break Lorentz reciprocity [37,38]. To calculate the diagonal term $\varepsilon_d$, we apply the Kubo Greenwood formalism within the random phase approximation to a two-band model with spin degeneracy. This formalism considers both interband and intraband transitions [48, 49]

$$\varepsilon_d = \varepsilon_b + i\frac{\sigma}{\omega} \quad (4)$$

σ is the bulk conductivity given by

$$\sigma = \frac{r_s g}{6}\Omega G\left(\frac{\Omega}{2}\right) + i\frac{r_s g}{6\pi}\left\{ \frac{4}{\Omega}\left[1 + \frac{\pi^2}{3}\left(\frac{k_B T}{E_F T}\right)^2\right] + 8\Omega \int_0^{\xi_C} \frac{G(\xi) - G\left(\frac{\Omega}{2}\right)}{\Omega^2 - 4\xi^2}\xi d\xi \right\} \quad (5)$$

$\varepsilon_b$ is the background permittivity, $\Omega = \hbar(\omega + i\tau^{-1})/E_F$ is the complex frequency normalized by the chemical potential, $\tau^{-1}$ is the scattering rate corresponding to Drude damping, $G(E) = n(-E) - n(E)$ where $n(E)$ is the Fermi distribution function, $E_F(T)$ is the chemical potential, $r_s = e^2/4\pi\varepsilon_0\hbar v_F$ is the effective fine structure constant, $v_F$ is the Fermi velocity, $g$ is the number of Weyl points, $\xi_c = E_c/E_F$ where $E_c$ is the cutoff energy beyond which the band dispersion is nonlinear. In this work, following Refs. [49, 50] we use the parameters $\varepsilon_b = 6.2$, $\xi_c = 3$, $\tau = 1000$ fs, $g = 2$, and $v_F = 0.83 \times 10^9$ m/s. The chemical potential as a function of temperature can be calculated from charge conservation [36, 39]

$$E_F(T) = \frac{2^{1/3}\left[9E_F(0)^3 + \sqrt{81E_F(0)^6 + 12\pi^6 k_B^6 T^6}\right]^{2/3} - 2\pi^2 3^{1/3} k_B^2 T^2}{6^{2/3}\left[9E_F(0)^3 + \sqrt{81E_F(0)^6 + 12\pi^6 k_B^6 T^6}\right]^{1/3}} \quad (6)$$

where $E_F$ (0 K) = 0.163 eV and $E_F$ (300 K) = 0.150 eV.

Figure 2 shows the real part of $\varepsilon_d$ and $\varepsilon_a$ at different momentum-separations $b$. The value of $\varepsilon_a$ heavily depends on momentum-separations $b$, and the magnitude of $\varepsilon_a$ can be comparable to that of $\varepsilon_d$ in the mid-infrared wavelengths, which indicates a highly unusual and extremely large gyrotropic response [39].

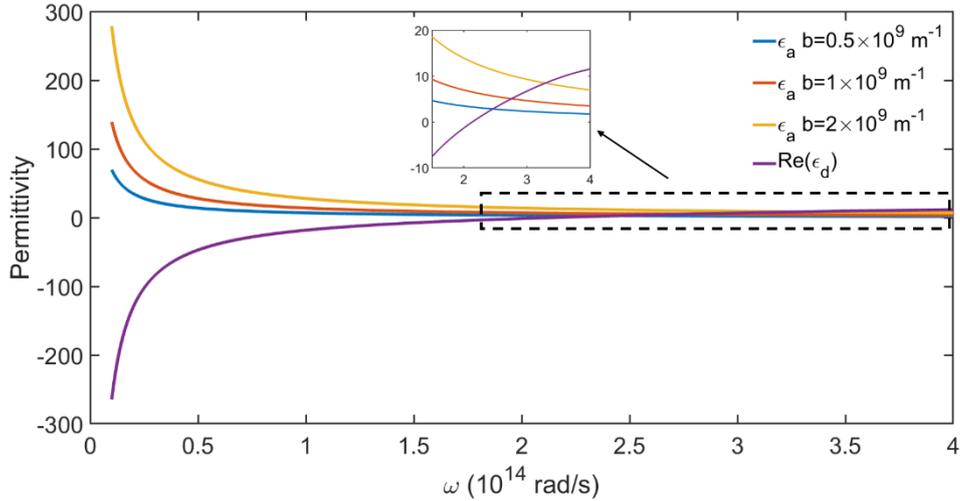

Fig. 2 Real part of permittivity $\varepsilon_d$ and $\varepsilon_a$ at different momentum-separations $b$, $T$ = 300 K.

When omitting the thermal emission to the background, the total heat power received by particle $i$ through vacuum can be written as [40]:

$$P_i = \int_0^\infty \frac{d\omega}{2\pi} P_{i,\omega} = 3\int_0^\infty \frac{d\omega}{2\pi}[\Theta(T_j) - \Theta(T_i)]\tau_{ji} \quad (7)$$

where $\Theta(T) = \hbar\omega/[\exp(\hbar\omega/k_B T) - 1]$ is the mean energy of the Planck thermal harmonic oscillators without zero-point energy. $\hbar$ is the reduced Planck constant and $k_B$ is the Boltzmann constant. $\tau_{ji} = 4k_0^4 \text{Tr}[\boldsymbol{\chi}_i \mathbf{G}_{ij} \boldsymbol{\chi}_j \mathbf{G}_{ij}^+]/3$ is the transmission coefficients for the power exchanged between the NPs, which reflects the number of modes in the radiative system [41, 42]. $\boldsymbol{\chi}_i = (\alpha_i - \alpha_i^\dagger)/2i$ is the response function of particle $i$ when considering fluctuation-electrodynamic theorem and avoiding unphysical effects [43, 44]. For an anisotropic particle, based on the Clausius-Mossoti form, when neglecting radiative correction, its polarizability as a function of permittivity tensor is frequency-dependent, orientation-dependent and temperature-dependent for WSM NPs [45, 46]. Polarizability consists of electrical polarizability and magnetic polarizability. For particles smaller than the wavelength, the contribution of magnetic moments can be neglected compared to the contributions of electric field in the dissipation process. The electric polarizabilities of the anisotropic NPs $\boldsymbol{\alpha}_i$ is given as

$$\boldsymbol{\alpha}_i = 4\pi R^3 \frac{\boldsymbol{\varepsilon}_i - \mathbf{1}\mathbf{I}}{\boldsymbol{\varepsilon}_i + 2\mathbf{I}} \tag{8}$$

$\mathbf{I}$ is the third order identity matrix, $\boldsymbol{\varepsilon}_1$ ($\boldsymbol{\varepsilon}_2$) is the permittivity tensor of emitter (receiver). For $\boldsymbol{\varepsilon}$ given in eq. 2, $\boldsymbol{\alpha}$ adopts the form

$$\boldsymbol{\alpha}_i = \begin{pmatrix} \alpha_{xxi} & 0 & 0 \\ 0 & \alpha_{yyi} & i\alpha_{yzi} \\ 0 & -i\alpha_{zyi} & \alpha_{zzi} \end{pmatrix} \tag{9}$$

in this case $\alpha_{yyi} = \alpha_{zzi}$. $\mathbf{G}_{ij}$ in $\tau_{ji}$ denotes the dyadic Green's tensor of the system, which can be written as

$$\mathbf{G}_{ij} = \mathbf{M}_{ij}^{-1} \mathbf{G}_{ij}^{(0)} \tag{10}$$

where $\mathbf{M}_{ij} = \mathbf{I} - k_0^4 \alpha_i \alpha_j \mathbf{G}_{ij}^{(0)} \mathbf{G}_{ij}^{(0)T}$ represents the multiple reflections between two particles [42]. $\mathbf{G}_{ij}^{(0)}$ is vacuum contribution to the Green's function, which is related to the location of particles [43]:

$$\mathbf{G}_{ij}^{(0)} = \mathbf{G}^{(0)}(\mathbf{r}_i, \mathbf{r}_j) = \frac{e^{ik_0 r_{ij}}}{4\pi r_{ij}} \left[ \left(1 + \frac{ik_0 r_{ij} - 1}{k_0^2 r_{ij}^2}\right) \mathbf{I} + \frac{3 - 3ik_0 r_{ij} - k_0^2 r_{ij}^2}{k_0^2 r_{ij}^2} \hat{\mathbf{r}}_{ij} \otimes \hat{\mathbf{r}}_{ij} \right] \quad (11)$$

in which $k_0$ is wave vector in vacuum and $r_{ij}=|\mathbf{r}_{ij}|$ is the magnitude of the vector linking position $\mathbf{r}_i$ and $\mathbf{r}_j$, $\hat{\mathbf{r}}_{ij} = \mathbf{r}_{ij} / r_{ij}$ and $\otimes$ denotes the outer product.

The radiative heat transfer between two blackbody spheres is considered for comparison. The total heat flux between two blackbody spheres is calculated in the framework of the traditional radiation transfer theory [39]

$$P_{ij} = \sigma_s \left(T_i^4 - T_j^4\right) A F_{ij} \quad (12)$$

where $\sigma_s$ is the Stefan-Boltzmann constant, and $A = 4\pi R^2$. $F_{ij}$ is the radiative view factor between the two spheres of equal radius, which is calculated from

$$F_{ij} = \frac{1}{2} \left\{ 1 - \left[1 - \frac{1}{\left(\frac{d-2R}{R}+2\right)^2}\right]^{1/2} \right\} \quad (13)$$

### 3. Results and discussion

Figures 3(a) and 3(b) show the total heat power between two blackbody NPs and WSM NPs at various temperatures. To clearly analyze the NFRHT between two WSM NPs, we firstly assume that the value of momentum-separation **b** is zero. As shown in Fig. 3(a), when the temperature of the emitter is fixed, the heat power always decreases monotonically with the increase of the temperature of the receiver, showing positive differential thermal conductance phenomenon. However, as shown in Fig. 3(b), when the temperature of the emitter is fixed, the heat flux does not decrease monotonically with the increase of the temperature of the receiver, showing NDTC phenomenon. Specifically, when the temperature of the emitter is 300 K, as shown by the black dashed line, the heat flux firstly increases and then decreases with the increase of the temperature of the receiver. In addition, it is clear the heat flux between two WSM NPs is higher than that between two blackbody NPs when the temperature of the emitter is about 300 K ~500 K. This is attributed to the excitation of localized surface plasmon modes, which will be revealed in the following.

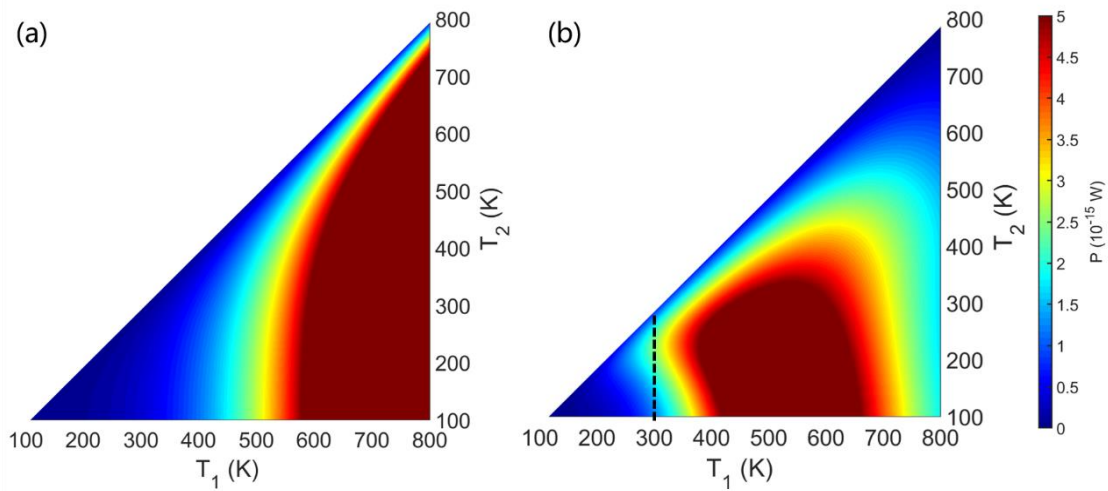

Fig. 3 The total heat power between two (a) blackbody, and (b) WSM NPs varying with temperature of emitter and receiver.

To reveal the mechanism of the NDTC effect in the NFRHT between two WSM NPs, the heat flux as a function of the temperature of the receiver is shown in Fig. 4 when the temperature of the emitter is fixed at 300 K. When the temperature of the receiver is higher than 220 K, the heat flux decreases with the increase of the temperature of the receiver. However, when the temperature of the receiver is lower than 220 K, the heat flux increases with the increase of the temperature of the receiver. According to Eq. (7), the heat flux is mainly influenced by the difference between the Planck oscillators of the emitter and receiver ($\Theta(T_1)-\Theta(T_2)$) as well as transmission coefficient. As shown in Fig. 4, the former decreases monotonically with the increases of the temperature of the receiver, while the latter does change monotonically with the temperature of the receiver. Particularly, when the temperature of the receiver is lower than 220 K, the transmission coefficient increases monotonically with the increases of the temperature of the receiver. Therefore, the NDTC effect mainly depends on the transmission coefficient. Besides, the heat flux between two blackbody NPs is also shown in the same figure for comparison. It is clear the heat flux decreases monotonically with the increases with the temperature of the receiver in this case.

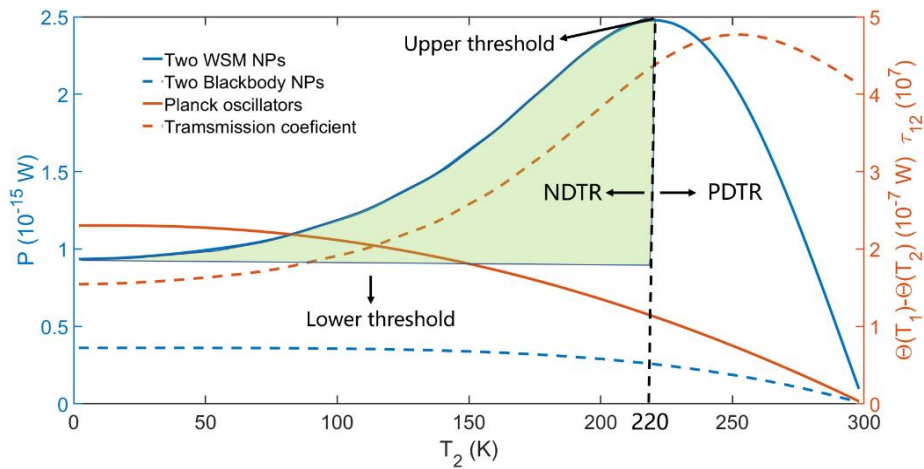

Fig. 4 The total heat power between WSM NPs, Planck oscillator, total heat power between blackbody NPs, and transmission coefficient between WSM NPs. The blue lines correspond to the left *y*-axis coordinate and the orange lines correspond to the right *y*-axis coordinate. The temperature of emitter $T_1$ is 300 K.

According to the analysis in Section 2, the transmission coefficient is related with the polarizability of the NPs, and it is affected by the permittivity. According to Eq. (8), the localized plasmon modes can be excited when $\varepsilon_d+2=0$ is satisfied [47]. Fig. 5 shows the real part of $\varepsilon_d$ varying with the angular frequency and temperature. The blue dashed line indicated the excitation of localized plasmon modes. It is clear the resonant frequency shifts to higher frequency with the decrease of the temperature.

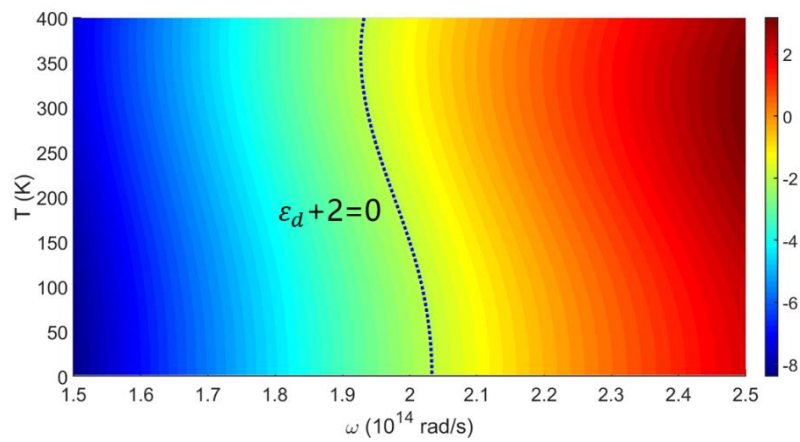

Fig. 5 Real part of $\varepsilon_d$ varying with angular frequency and temperature, the blue dotted line corresponds to the dispersion relation of nanoparticles.

When the temperature of the emitter is fixed at 300 K, spectral heat flux is shown in Fig. 6(a) for different temperatures of the receiver. When the temperature of the receiver is 220 K, the resonant frequency is $1.964 \times 10^{14}$ rad/s, and the spectral heat flux at this frequency is greatly enhanced. When the temperature of the receiver is 50 K and 280 K, the corresponding resonant frequencies are $2.029 \times 10^{14}$ rad/s and $1.939 \times 10^{14}$ rad/s, respectively. However, the spectral heat flux at resonant frequency is smaller, compared with the case when the temperature of the receiver is 220 K.

The spectral heat flux reflects the coupling between the polarizabilities of the emitter and receiver. As shown in Fig. 6(b), the polarizability at 300 K is enhanced at angular frequency of $1.935 \times 10^{14}$ rad/s, which is close to the resonant frequency at 280 K. However, because the amplitudes of the two polarizabilities are not high, spectral heat flux is not greatly enhanced. When the temperature is 50 K, polarizability is enhanced at angular frequency of $2.029 \times 10^{14}$ rad/s, and the amplitude of the polarizability is very larger. Nevertheless, the coupling between the polarizabilities at 50 K and 300 K is not very strong because the two resonant frequencies are far away. When the temperature is 220 K, polarizability is enhanced at angular frequency of $1.967 \times 10^{14}$ rad/s. With moderate amplitude of the polarizabilities, and moderate distance between two resonant frequencies, the coupling between the polarizabilities at 50 K and 300 K is the strongest. Therefore, the polarizability can well explain the NDTC effect. It is noted that the resonant frequencies at different temperatures agree very well with the theoretical analysis shown in Fig. 5.

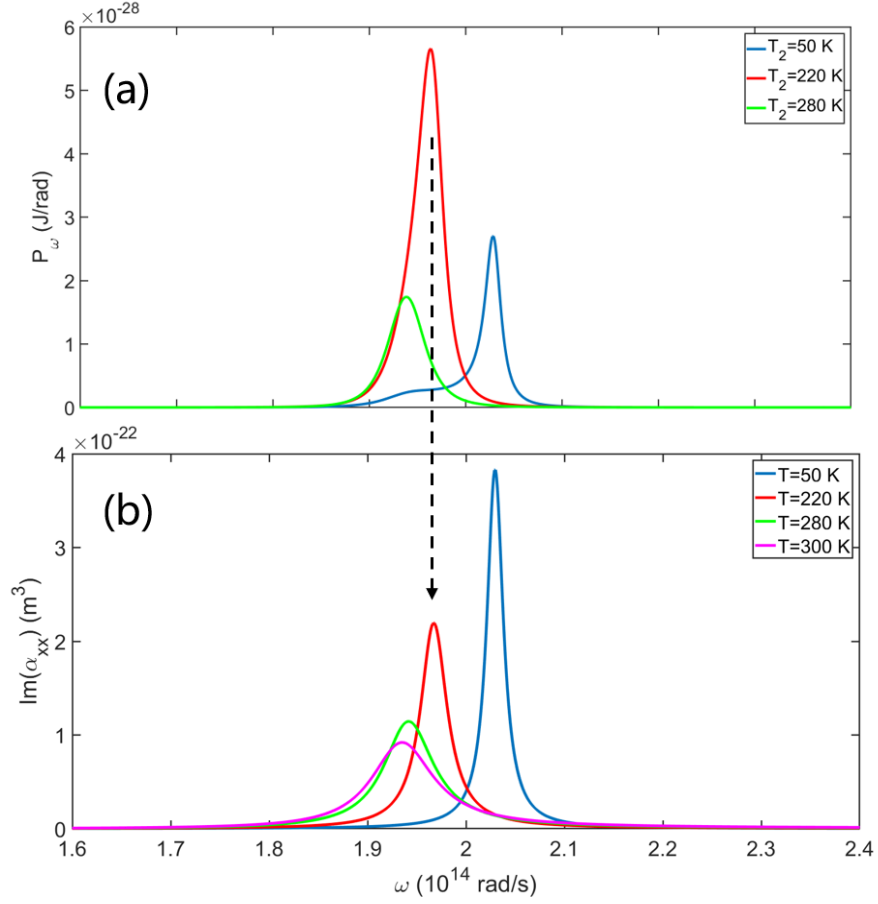

Fig. 6(a) Spectral heat power between WSM NPs of different temperature $T_2$ of receiver when $T_1$=300 K, and (b) imaginary part of $α_{xx}$ at different temperature.

Until now, we have only considered the case that the value of *b* is zero. As shown in Fig. 7, the heat flux as a function of the temperature of the receiver for different *b* is plotted. The NDTC effect always exists, regardless of the value of *b*. Besides, one can see that the heat flux can be further enhanced when *b* is $0.5 × 10^9$ m$^{-1}$ or $1 × 10^9$ m$^{-1}$. In addition, the heat flux is not always increasing with the increase of *b*, and it can be reduced when *b* is $2 × 10^9$ m$^{-1}$.

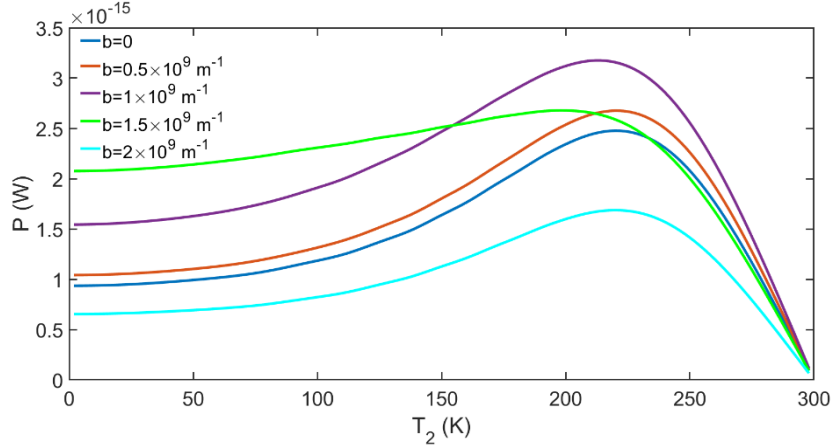

Fig. 7 The total heat power between WSM NPs varying with different temperature of receiver at different momentum-separation $b$.

To reveal the influence of the $b$ on the NDTC effect, the spectral heat flux is shown in Fig. 8(a) for different $b$. The temperature of the emitter and receiver is 300 K and 280 K, respectively. When $b$ is zero, there is only one peak at angular frequency of $1.964 \times 10^{14}$ rad/s, which is termed as Peak 2 and it is attributed to the excitation of localized plasmon modes. When $b$ is not zero, the peak at angular frequency of $1.964 \times 10^{14}$ rad/s still exists, and its amplitude is reduced. Besides, two peaks are generated at the left and right sides of angular frequency of $1.964 \times 10^{14}$ rad/s, which are termed as Peak1 and Peak 3, respectively. As $b$ increases, the Peak 1 tend to be excited at smaller frequency, while Peak 3 tend to be excited at larger frequency.

To find the mechanism for the Peak 1 and Peak 3, the polarizability component $\alpha_{yz}$ at temperature of 300 K is shown in Fig. 8(b). When $b$ is not equal to 0, the polarizability presents two peaks, which originates from local circular modes, leading to the enhanced spectral heat flux at Peak 1 and Peak 3 [35, 47, 48]. Since Fig. 8(a) shows the coupled results of the polarizability of two particles at different temperatures, while Fig. 8(b) shows the polarizability of a single particle. Thus, the angular frequencies of the peaks and dips are slightly different from that of the peaks of the radiative spectrum in Fig. 8(a). Through analyzing the polarizability of Clausius-Mossoti form, the analytical formula for resonance mode can be described as [47-51]:

$$\det(\boldsymbol{\varepsilon}(\omega,T)+2\mathbf{I})=0=(\varepsilon_d+2)\left((\varepsilon_d+2)^2-\varepsilon_a^2\right) \qquad (14)$$

where $T$ is a constant with a value of 300 K. According to Eq. (14), it can be seen that the resonance modes appear when $(\varepsilon_d+2)=0$ or $(\varepsilon_d+2)^2-\varepsilon_a^2=0$ is satisfied. The former corresponds to the localized plasmon modes, while the latter corresponds to the local circular modes, which agrees very well with our numerical results. Back to Fig. 8(b), one can see that with the increase of $b$, the absolute value of the polarizability corresponding to peak 1 and peak 3 decreases gradually, and the polarizability only manifests one peak when $b=2\times10^9\,\mathrm{m^{-1}}$. Besides, the difference between the frequencies of peak 1 and peak 3 increases with the increase of $b$. This phenomenon is attributed to the gyrotropic response of WSM NPs. The modal fields corresponding to Peak 1 and Peak 3 rotate clockwise and counterclockwise along the momentum-separation $b$, respectively.

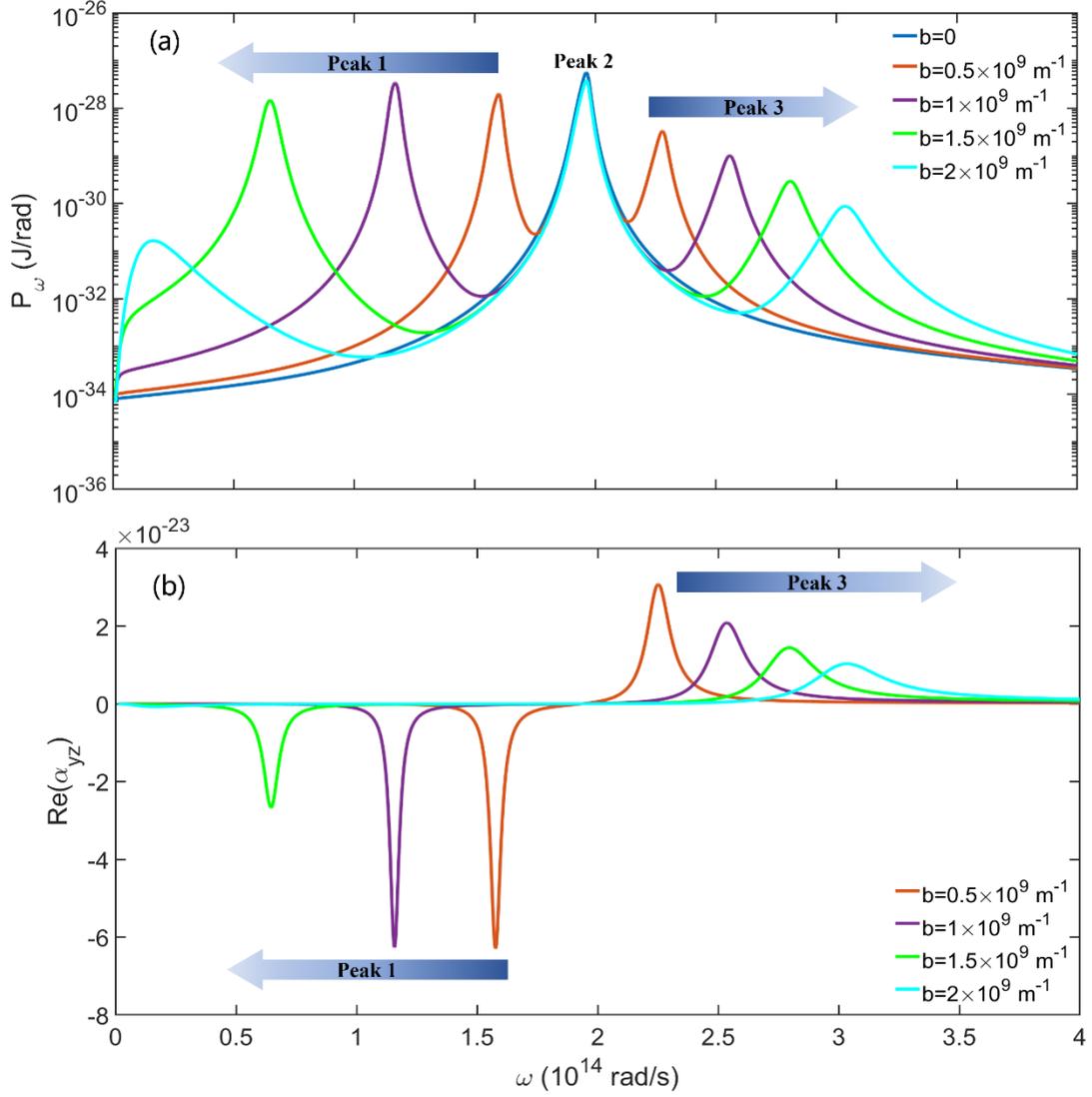

Fig. 8 (a) Spectral heat power between two WSM NPs of different momentum-separation $b$ when $T_1 = 300$ K and $T_2 = 220$ K, and (b) Real part of $\alpha_{yz}$ at different momentum-separation $b$ when temperature of NP is 300 K.

Taking the momentum-separation $1\times10^9$ m$^{-1}$ for example, we discuss the transmission coefficient under different temperatures, as shown in Fig. 9. Clearly, there are three thin bright band, which correspond to Peak 1, Peak 2 and Peak 3 in the Fig. 8. It can be found that bright band is most evident at Peak 2, especially at higher temperatures, indicating that the localized plasmon mode makes the main contribution to the large heat flux. However, the transmission coefficients at Peak 1 and Peak 3 are relatively small. Thus, the contribution of the local circular modes to the enhanced heat flux is relatively weak. The dashed lines in Fig. 9 show the solution of Eq. 14, which can be in good agreement with the bright bands. The results further illustrate that the localized plasmon mode and local circular mode can enhance theNFRHT.

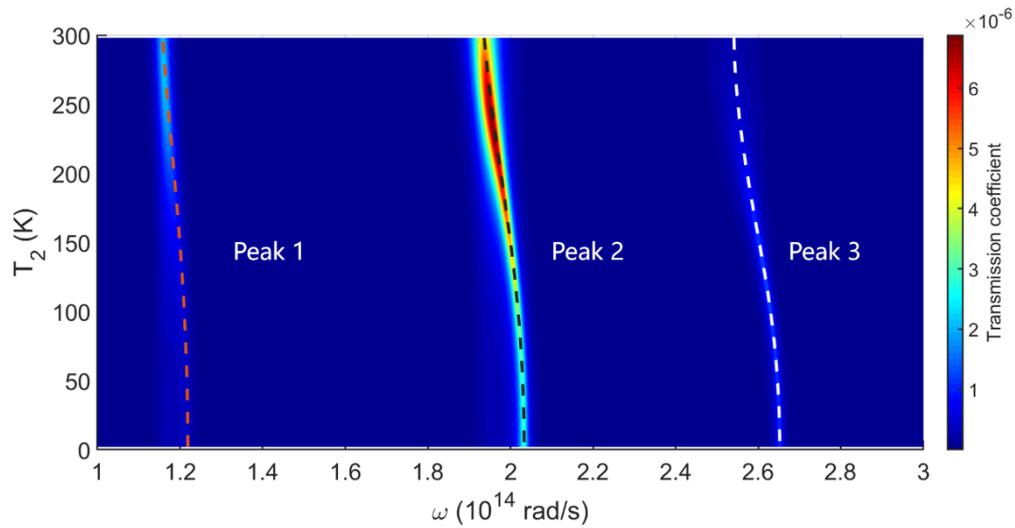

Fig. 9 The transmission coefficient between WSM NPs distribution in frequency and temperature. $b = 1\times10^9$ m$^{-1}$, $T_1 = 300$ K. The pink, black, and white dashed lines correspond to the dispersion curves of Peak 1, Peak 2 and Peak 3, respectively.

### 4. Conclusions

In summary, we have investigated the NFRHT between two Weyl semimetal nanoparticles. The numerical results show that the heat flux does not change monotonically with the temperature of receiver at a certain temperature of emitter, which is the NDTC effect. The key factor for the NDTC phenomenon is the temperature-dependent permittivity of WSM. By analyzing the coupling of particle polarizability at different temperatures, the physical mechanism of NDTC phenomenon is revealed. It is found that the NDTC effct is contributed to the localized plasmon mode and the localized circular modes, and the NFRHT can be significantly enhanced by the localized plasmon mode. This work has potential significance in the regulation of NFRHT.

**Acknowledgments**

This work supported by the National Natural Science Foundation of China (Grant Nos. 52106099, and 51976173), the Shandong Provincial Natural Science Foundation (Grant No. ZR2020LLZ004), the Jiangsu Provincial Natural Science Foundation (Grant No. BK20201204), the Basic Research Program of Taicang (Grant No. TC2019JC01), and the Fundamental Research Funds for the Central Universities (Grant No. D5000210779).